\documentclass[amssymb,amsmath,aip,reprint,]{revtex4-1}

\usepackage{graphicx}
\usepackage{dcolumn}
\usepackage{bm}

\usepackage{float}
\usepackage[utf8]{inputenc}
\usepackage[T1]{fontenc}
\usepackage{mathptmx}
\usepackage{etoolbox}
\usepackage{xcolor}
\usepackage{SIunits}
\usepackage[textsize=tiny]{todonotes}
\usepackage{chngcntr}

\usepackage[final]{changes}

\makeatletter
\def\@email#1#2{%
 \endgroup
 \patchcmd{\titleblock@produce}
  {\frontmatter@RRAPformat}
  {\frontmatter@RRAPformat{\produce@RRAP{*#1\href{mailto:#2}{#2}}}\frontmatter@RRAPformat}
  {}{}
}%
\makeatother
\begin{document}

\newcommand{\fg}{\textcolor{red} }
\newcommand{\tf}{\textcolor{blue} }
\newcommand{\vr}{\textcolor{teal} }
\newcommand{\lb}{\textcolor{orange} }
\newcommand{\sa}{\textcolor{green} }

\newcommand{\luigi}[2][]{\todo[color=orange!20,#1]{{\bf LB:} #2}}
\newcommand{\valerio}[2][]{\todo[color=teal!20,#1]{{\bf VR:} #2}}
\newcommand{\thorben}[2][]{\todo[color=blue!20,#1]{{\bf TF:} #2}}
\newcommand{\francesco}[2][]{\todo[color=red!20,#1]{{\bf FG:} #2}}
\newcommand{\simone}[2][]{\todo[color=green!20,#1]{{\bf SA:} #2}}

\preprint{AIP/123-QED}

\title{DeepLNE++ leveraging knowledge distillation for accelerated multi-state path-like collective variables}

\author{Thorben Fr\"ohlking}%
\affiliation{School of Pharmaceutical Sciences, University of Geneva, Rue Michel Servet 1, 1206, Genève, Switzerland}
\affiliation{Institute of Pharmaceutical Sciences of Western Switzerland (ISPSO), University of Geneva, 1206, Genève, Switzerland}
\affiliation{Swiss Institute of Bioinformatics, University of Geneva, 1206, Genève, Switzerland}


\author{Valerio Rizzi}%
\affiliation{School of Pharmaceutical Sciences, University of Geneva, Rue Michel Servet 1, 1206, Genève, Switzerland}
\affiliation{Institute of Pharmaceutical Sciences of Western Switzerland (ISPSO), University of Geneva, 1206, Genève, Switzerland}
\affiliation{Swiss Institute of Bioinformatics, University of Geneva, 1206, Genève, Switzerland}

\author{Simone Aureli}%
\affiliation{School of Pharmaceutical Sciences, University of Geneva, Rue Michel Servet 1, 1206, Genève, Switzerland}
\affiliation{Institute of Pharmaceutical Sciences of Western Switzerland (ISPSO), University of Geneva, 1206, Genève, Switzerland}
\affiliation{Swiss Institute of Bioinformatics, University of Geneva, 1206, Genève, Switzerland}

\author{Francesco Luigi Gervasio}
\email{francesco.gervasio@unige.ch}
\affiliation{School of Pharmaceutical Sciences, University of Geneva, Rue Michel Servet 1, 1206, Genève, Switzerland}
\affiliation{Institute of Pharmaceutical Sciences of Western Switzerland (ISPSO), University of Geneva, 1206, Genève, Switzerland}
\affiliation{Swiss Institute of Bioinformatics, University of Geneva, 1206, Genève, Switzerland}
\affiliation{Department of Chemistry, University College London, London, WC1E 6BT, United Kingdom}


\date{\today}

\begin{abstract}

Path-like collective variables can be very effective for accurately modeling complex biomolecular processes in molecular dynamics simulations. Recently, we introduced DeepLNE, a machine learning-based path-like CV that provides a progression variable $s$ along the path as a non-linear combination of several descriptors, effectively approximating the reaction coordinate. However, DeepLNE is computationally expensive for realistic systems needing many descriptors and limited in its ability to handle multi-state reactions. Here we present DeepLNE++, which uses a knowledge distillation approach to significantly accelerate the evaluation of DeepLNE, making it feasible to compute free energy landscapes for large and complex biomolecular systems. In addition, DeepLNE++ encodes system-specific knowledge within a supervised multitasking framework, enhancing its versatility and effectiveness. 

\end{abstract}

\maketitle
Classical molecular dynamics (MD) simulations at the atomistic scale provide a powerful method for studying complex physical, chemical, and biological systems.
Indeed, MD simulations can reveal mechanisms at spatial and temporal scales that are difficult to observe experimentally, acting as a computational microscope.\cite{Shaw2012} Advances in computational power and force field accuracy have increased the capabilities of MD, even though a full description of intricate biological phenomena remains hard to achieve due to limitations in accessible timescales. To address this, enhanced sampling algorithms have been developed to allow the exploration of high-energy transitions and their associated free energy landscapes.~\cite{Laio2008,Valsson2016,Camilloni2018} Many of these algorithms use either collective variables (CVs) or paths connecting two-states A and B, each of which has its own advantages and limitations.~\cite{Laio2002,Bolhuis2002}

CV-based algorithms rely on the identification of slow degrees of freedom specific to the system under investigation and are limited to applying an external bias potential only on the empirically determined relevant ones.~\cite{Pietrucci2017} Path-based methods, on the other hand, can incorporate many degrees of freedom but are highly dependent on the definition of end states and effective path search algorithms. Path-collective variables (PATHCVs) combine these approaches and are often coupled with enhanced sampling algorithms to explore high free energy barriers and successfully reconstruct free energy profiles for complex systems.~\cite{Branduardi2007}

It is worth emphasizing that the effectiveness of PATHCVs depends on the metric and milestone parameters chosen. Recent work has explored machine learning to optimise PATHCV metrics, but linear combinations of features can be sub-optimal for complex pathways, such as ligand binding to a protein.~\cite{Hovan2019} Non-linear combinations of descriptors better capture the relevant degrees of freedom at different stages of the pathway.~\cite{Bonati2020,Bonati2023} Data-driven methods have also emerged for learning CVs, using techniques such as variational autoencoders and kernel ridge regression.~\cite{Sipka2023,Pietrucci2024}

We recently introduced a new machine learning approach, DeepLNE (deep-locally non-linear-embedding), which refines the PATHCV method and overcomes existing limitations \cite{Froehlking2024}. DeepLNE creates a semi-automatic 1D description of the training data by combining local linear embedding~\cite{Roweis2000} (LLE) principles with PATHCV. The algorithm uses dimensionality reduction via artificial neural networks (ANNs), differentiable k-nearest neighbour selection, and an ANN-encoded 1D latent space. By focusing on system dynamics, DeepLNE forms a path-like CV anchored by training data, expanding the tools for CV definition and improving on previous methods with respect to available input space and functional flexibility.

The algorithm can be applied to reactive MD trajectories connecting metastable states and derives a PyTorch~\cite{Paszke2019} model that accurately approximates the true reaction coordinate of the system $s$ and its perpendicular distance $z$. Therefore, exporting DeepLNE as a CV and biasing it via state-of-the-art enhanced sampling methods such as 'On-the-fly Probability Enhanced Sampling' (OPES)~\cite{Invernizzi2020,Invernizzi2022} or OneOPES~\cite{Rizzi2023} allows efficient estimation of the free energy surface (FES).
However, DeepLNE is computationally demanding. When the underlying feature space is large and models are anchored to many training datapoints, the model evaluation via PyTorch can significantly impact the computational performance of the MD engine. Additionally, when multiple metastable states are aligned along the reaction path, the sequentiality of assigned path values on the 1D projection automatically derived by the DeepLNE algorithm can be different from the desired one.
In this study, we show how DeepLNE can be improved by defining a multi-tasking objective function aligning the expectations when designing the path-like CVs (e.g. sequentiality of metastable states) with the model output and how the evaluation of DeepLNE CVs can be significantly accelerated by representing and exporting them as artificial neural networks (ANN). We achieve this within a knowledge distillation framework deriving DeepLNE student models for the path-like variables $s$ and $z$. 
In the paper, we also document the best practices for training DeepLNE models including cross-validation and augmented objective functions allowing the use of labeled data in a supervised multi-tasking framework as well as the option to assign importance weights to individual input features.
We demonstrate the extended and improved algorithm on the toy models: 3-state M\"uller-Brown-potential, alanine tripeptide, and the microRNA precursor pre-miR21. Importantly we benchmark the original PyTorch-based CV against the computationally faster DeepLNE student models obtained from knowledge distillation, showing that for the solvated pre-miR21 system composed of 31 nucleotides with a total of 45543 atoms we can achieve a 3 times speed-up.

\section{Methods}


Before introducing the DeepLNE++ algorithm and its novelties, we summarise the original DeepLNE framework, highlighting its advantages and its limitations.

\subsection{DeepLNE}

The DeepLNE CV aims to represent high-dimensional datasets using a single dimension. It combines ideas from PATHCV and LLE within a neural network architecture which parameters are optimized using the mean square error between the original input vector $\bm{X}$ and the reconstructed one $\bm{\hat{X}}$ for $m$ training datapoints:
\begin{equation}
    \mathcal{L}=\frac{1}{m}\sum_{i=1} ^m | \bm{X}-\bm{\hat{X}} |^2
\end{equation}
The architecture can be seen as a generalized autoencoder.
An initial ANN reduces dimensionality. Then a continuous k-nearest-neighbor (k-NN) step is applied to each data point.
The neighborhood information is compressed into a one-dimensional representation $s$. This representation is used for reconstruction via a decoder and for computing an accessory perpendicular distance $z$
Since $s$ is derived from the features of a data point’s neighbors it is robust to extrapolation. However, evaluating the model becomes slower with larger datasets due to memory requirements. The DeepLNE algorithm constructs variables $s$ and $z$ that approximate a transition path between states. It addresses several issues encountered in PATHCVs:
\begin{itemize}
    \item Automatic Metric Learning ($d$): DeepLNE learns a lower-dimensional metric to reduce degeneracy when identifying local neighborhoods.
    \item Automatic Neighborhood Construction: Neighborhoods are constructed using a differentiable k-NN step.
    \item Heterogeneous Features: Distances, angles, and contact maps can be combined in $\bm{X}$ without predefined landmarks.
\end{itemize}

However, there are remaining limitations that we attempted to address in this study:
\begin{itemize}
    \item Sequentiality Preservation: The 1D DeepLNE variable $s$ may not preserve desired sequentiality in multi-state systems.
    \item Computationally Demanding Evaluation: On-the-fly evaluation for biasing MD simulations can impact performance.
\end{itemize}

\subsection{DeepLNE++ in Knowledge Distillation framework}

\begin{figure*}[htb]
\includegraphics[width=0.6\textwidth]{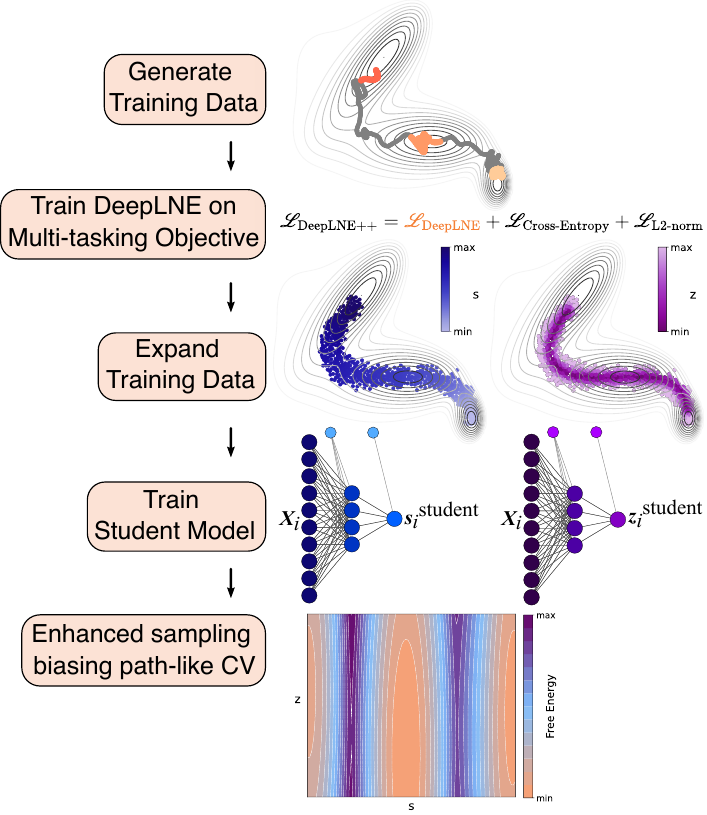}
\caption{\label{fig:SchematicCV} Schematic of the steps within the knowledge distillation framework for DeepLNE++. The starting point is the generation of training datapoints to which labels are assigned to \textit{a priori} known metastable states along the transition path of a biophysical system, here represented by the trajectory of a particle in the M\"uller-Brown potential.
DeepLNE++ optimizes the model parameters using a Multi-tasking objective function $\mathcal{L}$. Its components are the DeepLNE reconstruction loss, a cross-entropy term with respect to the assigned labels and a L2-regularization term applied on the model parameters. We select the DeepLNE model that minimized the $\mathcal{L}$ loss evaluated on the validation data left-out from the training dataset and consider it as the teacher model. We use it to perform short simulations biasing its path-like variables $s$ and $z$ and thereby expand the original training dataset.
Next we train ANNs as DeepLNE student models to maximally reconstruct the teacher model output. We depict the weights and biases of the individual transformer components of the student models as colored circles. For these models we choose architectures with less free parameters than the teacher model in order to obtain models that can be evaluated at a lower computational cost. We train the DeepLNE student models via a reconstruction loss and choose the student models that minimize the loss evaluated on a validation dataset resulting in the most generalizing models. The ANN CVs obtained in this way are exported and used to bias MD simulations in enhanced sampling protocols. Reweighting the MD timeseries with respect to the path-like variables $s$ and $z$ allows estimating the underlying 2D FES of the system.
}
\end{figure*}

In the following, we describe all the key steps of our recommended strategy for training the DeepLNE++ CV (compare Fig.~\ref{fig:SchematicCV}).

\subsubsection{Training data generation}
We use 'ratchet-and-pawl' restraints of the Adiabatic Bias Molecular Dynamics (ABMD) method~\cite{Camilloni2011} to perform directed enhanced sampling simulations connecting all the relevant states. Here we choose the spring constant $\kappa$ such that we reach the target state within picoseconds of simulation time. $\kappa$ can be seen as a hyperparameter of the DeepLNE algorithm, because it regulates the directionality and the input feature space the DeepLNE model is trained on. In the limit of high $\kappa$, we connect all states of interest within a short simulation time. However, since the training data will be obtained from proportionally out-of-equilibrium dynamics, the minimum transition path has to be recovered \textit{a posteriori} from the trajectory after performing enhanced sampling simulation using the DeepLNE student models. We will show how this can be achieved when inspecting the FES with respect to the variables $s$ and $z$ in the following chapters. Consequently, we recommend not applying harmonic restraints on the $z$ variable as long as the minimum free energy path between the states of interest is unknown. Additionally, we recommend to generate ABMD states-connecting trajectories along all possible regions of the accessible descriptor space $\bm{X}$. The more exhaustive is the exploration of the input descriptor space provided during the training of DeepLNE, the less extrapolation is required and unexpected behavior can be minimized. A novel key step in the DeepLNE knowledge distillation framework is the assignment of labels to supervise the model. Information should be provided in the form of labels whenever a region in the input descriptor space can be assigned to a metastable state or a transition state with high confidence. To minimize the computational cost of DeepLNE we use Farthest Point-Sampling (FPS)\cite{Goscinski2023} to select a maximally diverse subset of the sampled data, such that only the $m$ frames are retained. The dataset is then split into training and validation data, with 80\% of the data used for training and the remaining 20\% is left out for validation.

\subsubsection{Teacher model training}
\label{train_teacher_model}
For all the DeepLNE teacher models in this study, we choose $d=3$ for the embedded manifold coordinates $\bm{x}_i$ and $k=3$ meaning we use 3 nearest-neighbors to construct the vector $\bm{x}_i^\mathrm{k-NN}$ (compare chapter of DeepLNE on hyperparameter choices in Ref.~\onlinecite{Froehlking2024}). We scale the $s$ variable using a sigmoid activation in the final encoder layer, restricting its values to the range of 0–1. As shown in Fig.~\ref{fig:SchematicCV} the network parameters are optimized using the mean square discrepancy between the original input and the decoded ones, also referred to as reconstruction loss term.
The model is trained on a multi-tasking objective function that includes a cross-entropy term allowing to supervise the model via labeled data and an L2-regularization term penalizing large magnitudes in the model parameters $p$:
\begin{align}
\label{eq:TrainingLossTeacher}
   \mathcal{L}_{train,~teach} = &\frac{1}{m} \sum_{i=1}^{m} \sum_{j=1}^{D} \alpha_j | \mathbf{X}_{ij} - \mathbf{\hat{X}}_{ij} |^2 \nonumber \\
   &+ \beta \left( -l_i \log(s_i) - (1 - l_i) \log(1 - s_i) \right) \nonumber \\
   &+ \gamma \sum_{j=1}^{P} p_j^2
\end{align}
Each term of the objective function has associated hyperparameters $\alpha$, $\beta$, and $\gamma$ that are found by scanning a range of different choices respectively. While $\alpha$ is a hyperparameter vector that tunes the importance of each individual input feature reconstruction, $\beta$ and $\gamma$ are scalar hyperparameters tuning the importance of the label reconstruction and the magnitude of the model parameters respectively.
In order to obtain the most generalizing model we select the one that minimizes the multi-tasking reconstruction loss on the validation data:
\begin{align}
    \mathcal{L}_{val,~teach} = &\frac{1}{m} \sum_{i=1}^{m} \sum_{j=1}^{D} \alpha_j | \mathbf{X}_{ij} - \mathbf{\hat{X}}_{ij} |^2 \nonumber \\
    &+ \beta \left( -l_i \log(s_i) - (1 - l_i) \log(1 - s_i) \right)
\end{align}
We use gradient descent using the ADAM optimizer with a learning rate of $10^{-3}$ and 5000 epochs, using the machine learning library PyTorch \cite{Paszke2019}.

\subsubsection{Expanding training data}

Since the student models will be more susceptible to overfitting, because the ANNs are not anchored to the training data as it is the case for the DeepLNE teacher model we recommend expanding the training data by performing short enhanced sampling simulations e.g. using OPES in exploration mode~\cite{Invernizzi2022} (OPES-EXPLORE) using the DeepLNE teacher model (in the order of ns). This step aims at increasing the robustness of the DeepLNE student models. To track the exploration of the descriptor space it can be instructive to plot the DeepLNE CVs along a few important physical descriptors or against the first principal components of the input features $\bm{X}$. Based on this inspection we empirically decide when the descriptor space has been explored sufficiently such that we are minimizing the regions in configurational space that need to be extrapolated by the DeepLNE student models. Ultimately the original training data are concatenated with the newly sampled datapoints. Also here we use FPS to select a maximally diverse subset of all available datapoints. We retain $m$ frames and we split the data into training and validation set at a 80-20 ratio.

\subsubsection{Student model training}
\label{train_student_model}
As shown in Fig.~\ref{fig:SchematicCV} we project the DeepLNE teacher model variables $s$ and $z$ on the extended training dataset and train DeepLNE student models by minimizing the mean square discrepancy between the teacher model output and the student model output:
\begin{align}
    \mathcal{L}_{train,~stud}=& \frac{1}{m}\sum_{i=1}^{m} \text{SmoothL1}(\xi_i^{\text{DeepLNE}} - \xi_i^{\text{Student}}) \nonumber \\
    &+ \gamma \sum_{j=1}^{P} p_j^2
\end{align}
with $\xi_i \in \{s_i, z_i\}$ and the PyTorch loss function SmoothL1 with $\beta=1.35$. The hyperparameter $\gamma$ regulates the penalty applied based on the magnitude of the student model parameters $p$. The choice of SmoothL1 loss is motivated by its robustness against outliers and consequently reducing overfitting in trained models. 
As the best model, we choose the one that minimizes this objective function evaluated on the validation dataset:
\begin{equation}
    \mathcal{L}_{\text{val,~stud}} = \frac{1}{m}\sum_{i=1}^{m} \text{SmoothL1}(\xi_i^{\text{DeepLNE}} - \xi_i^{\text{Student}})
\end{equation}
We use gradient descent using the ADAM optimizer with a learning rate of $10^{-3}$ and up to 15000 epochs, using the machine learning library PyTorch \cite{Paszke2019}. We recommend using a higher number of training epochs than in \ref{train_teacher_model} in order to obtain models with a lower validation error, which typically corresponds to models that overfit less and exhibit smooth extrapolation into unknown regions of the descriptor space $\bm{X}$.

\subsubsection{Exporting the student models into PLUMED}
The parameters of the trained DeepLNE student models are stored such that the models can be evaluated on-the-fly in MD simulations using the PLUMED plugin~\cite{Tribello2014}. We use a variation of the MULTI\_ANN CV available at \url{https://github.com/bussilab/plumed-multi-ann} that provides computationally efficient evaluation of ANN as CVs. An implementation of the full knowledge distillation framework and usage as a tutorial are available at \url{https://github.com/ThorbenF/DeepLNE2}.


\subsection{Interpretation of hyperparameters}

In order to obtain the DeepLNE models that are most adequate for the biophysical system at hand we here focus on the choice and the interpretation of the hyperparameters that are part of the knowledge distillation framework.

\subsubsection{Teacher model}

The teacher model keeps the original DeepLNE hyperparameter for the continuous k-nearest-neighbors step that are the number of neighbors $k$ and the sparsity of the selection matrix $t$, which detailed description can be found in Ref.~\onlinecite{Froehlking2024}. We recommend using $k=3$ and $t=0.1$ for models with high locality and numerical stability. The $z$ variable is also dependent on $\lambda$ which is not relevant for the training of the model and can be quickly adjusted without retraining the model. The effect of $\lambda$ is analogous to its role in the PATHCV. To decide the adequate value for all hyperparameters it can be helpful to visualize the resulting behavior of variables $s$ and $z$ in the descriptor space or the 2 first components of the PCA on the descriptor set (compare Ref.~\onlinecite{Froehlking2024}).

Beyond the DeepLNE hyperparameter in \ref{train_teacher_model} we introduced a new objective function for training the teacher model that contains hyperparameters $\alpha, \beta$ and $\gamma$, for which we have the following limiting cases:

\begin{itemize}
    \item $\alpha\rightarrow 0$: DeepLNE does not prioritize descriptor reconstruction.
    \item $\alpha\rightarrow 1$: Full reconstruction of the selected descriptor from $s$.
    \item $\beta\rightarrow 0$: Assigned labels are ignored.
    \item $\beta\rightarrow\infty$: Optimal matching of $s$ with all labels.
    \item $\gamma\rightarrow 0$: In this limit we obtain a model with maximal flexibility, which can lead to overfitting.
    \item $\gamma\rightarrow\infty$: Maximal penalty for large parameters, resulting in reduced model flexibility.
\end{itemize}

\subsubsection{Student model}

In \ref{train_student_model} we introduced a new objective function for the training of the student model that contains hyperparameters $\beta$ and $\gamma$. In contrast to the teacher model’s objective function, which uses the hyperparameter $\alpha$ to weight the importance of the input features, the student model’s objective function is missing $\alpha$. The motivation for this choice is that the student model can learn automatically and unsupervised which of the input descriptors $\bm{X}$ are relevant for reconstructing the DeepLNE teacher model variables $s$ and $z$.
While $\gamma$ behaves as just described above, we have the limiting cases for $\beta$:

\begin{itemize}
\item$\beta\rightarrow 0$: the SmoothL1 loss converges to L1 loss decreasing the weight of the discrepancies between teacher and student model on the objective function.
\item$\beta\rightarrow1$: the SmoothL1 objective becomes the L2 norm increasing the weight of the discrepancies between teacher and student model.
\end{itemize}

In this study, we assigned the same model specific hyperparameter choices to the training and the validation objective function.

\subsection{Details on MD simulations and DeepLNE training}

\begin{figure*}
\includegraphics[width=0.4\textwidth]{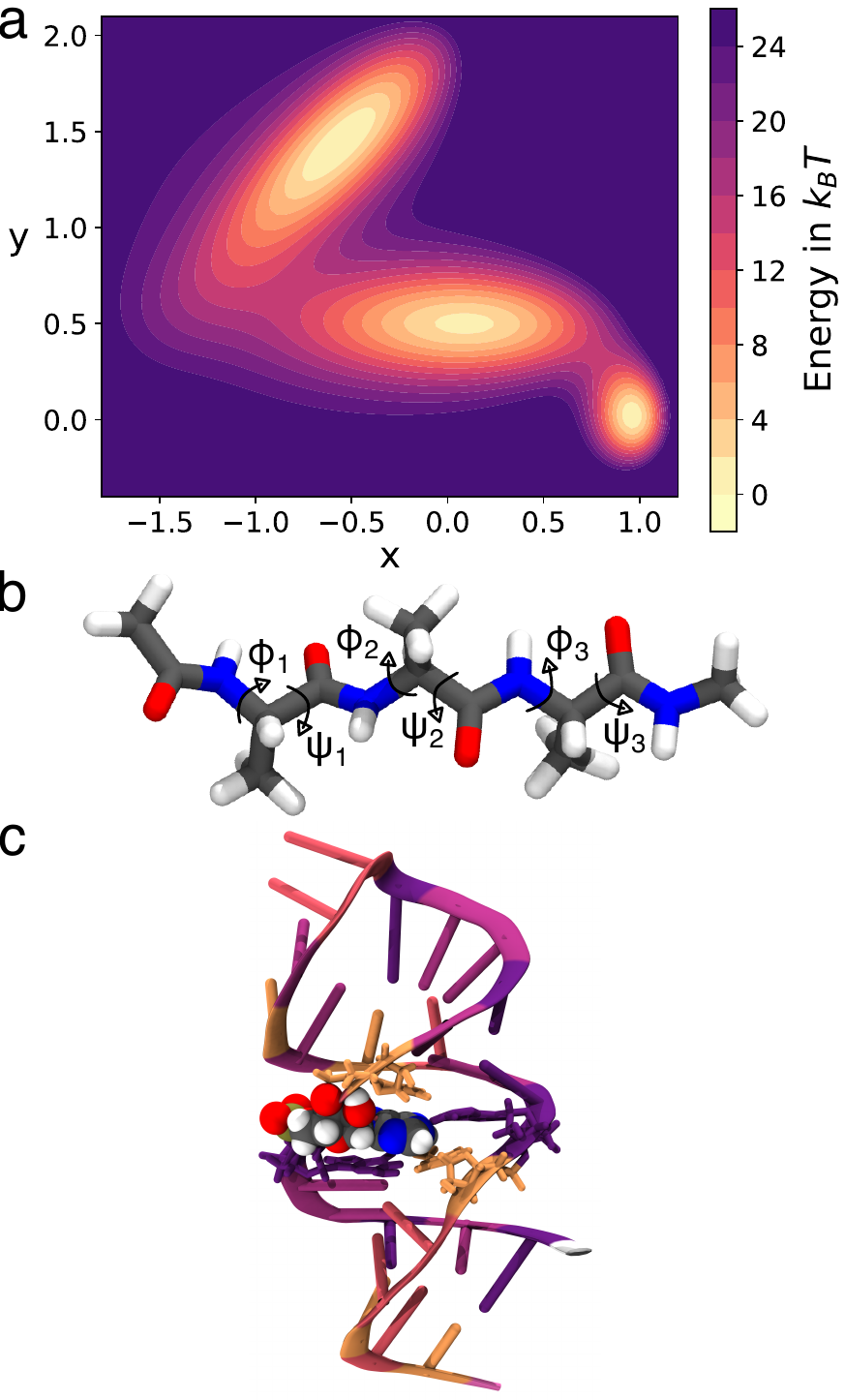}
\caption{\label{fig:SchematicSystems}
The DeepLNE CVs are tested on 3 toy models: (a) M\"uller-Brown potential used as a 3-state potential energy landscape for the simulation of a particle moving in two dimensions. (b) Structure of the well-studied biomolecule alanine tripeptide, a system that can be sufficiently described via its dihedral angles $\phi_1$, $\phi_2$, $\phi_3$ and $\psi_1$, $\psi_2$, $\psi_3$. (c) Structure of pre-miR21 composed of 31 nucleotides. The RNA is coloured with respect to the type of nucleotide: Guanine-purple, Adenine-light crimson, Cytosine-orange, Uracil-pink, G5-white, C3-mauve. We show a structure from G22 to C54 where the Adenine nucleobase A29 (highlighted via 'VDW' style) is integrated within the helical structure forming hydrogen bonds with the surrounding nucleobases (highlighted via 'licorice' style), rather than extending out towards the solvent.}
\end{figure*}

The DeepLNE knowledge distillation framework that we introduce here (see Fig.~\ref{fig:SchematicCV}) aims at representing the DeepLNE CVs $s$ and $z$ as a computationally cheap ANNs that can be straightforwardly exported into PLUMED. For all the systems investigated in this paper, we followed the same protocol: (1) sampling of the transition of interest; (2) training the original DeepLNE CV on a set of input features as a teacher model; (3) performing enhanced sampling MD simulation using the teacher model to extend the training data; (4) training a student model for $s$ and another student model for the $z$ variable; (5) exporting the CVs to PLUMED and combining them with an enhanced sampling method for free energy estimation. In the following, we outline all the details of our recommended strategy.
We test the DeepLNE knowledge distillation framework across the following applications: exploration of the 3-state M\"uller-Brown potential by a single particle, and sampling of the conformational space of the alanine tripeptide. Lastly, as an example of a real-world biological problem, we investigated the conformational change associated to the maturation of the microRNA's pre-miR21 (see Fig.~\ref{fig:SchematicSystems}), a prominent oncological target whose conformational space and interactions with a novel inhibitor has been recently studied in Ref.~\onlinecite{Aureli2024}. 

\subsubsection{M\"uller-Brown-Potential}

The simulations of a particle in the M\"uller-Brown potential are performed using the simple Langevin dynamics code contained in the VES module of PLUMED~\cite{Valsson2014}. The particle is moving in two dimensions under the action of the 3-state potential depicted in Fig.~\ref{fig:SchematicSystems} built out of a sum of Gaussians. We start by generating training data concatenating 2 ABMD simulations biasing the y-coordinate with a spring constant of $\kappa=30,80$~$k_BT$ respectively in order to connect the 3 metastable states. The DeepLNE CVs $s$ and $z$ are trained using the $x$ and $y$ position of the particle as input features $\bm{X}$ also assigning labels to the 3 metastable states. We train the DeepLNE model with hyperparameters $k=3$, $t=0.1$, $\beta=0.1$, $\gamma=1e^{-6}$ setting $\alpha=1$ for both input variables. The teacher model is then used to perform an exploration simulation to expand the training dataset. We deposit bias along the DeepLNE variable $s$ via the OPES-EXPLORE engine. We reduce the number of training datapoints to $m=1600$ applying FPS. The DeepLNE student models for $s$ and $z$ are subsequently trained on this expanded training dataset choosing $\gamma=1e^{-4}$. Finally we export the DeepLNE student models as CVs in PLUMED in order to perform free-energy estimation by depositing bias on $s$ via OPES and applying a half parabolic harmonic constraint on $z$ with a cutoff value of 0.1 (UWALL) and $\kappa = 2000~k_BT$. We set a bias deposition rate (PACE) of 200 steps and a barrier parameter equal to 15 kJ/mol. We estimate statistical errors on the FES via block analysis using 10 blocks.

\subsubsection{Alanine Tripeptide}

For the extensively studied FES of alanine tripeptide, $\phi_1$ and $\phi_2$ are commonly used to describe the metastable states of the system. However there are the additional dihedral angles $\phi_3$, $\psi_1$, $\psi_2$ and $\psi_3$ leading to at least 6 available dihedral angle descriptors for this system (compare Fig.~\ref{fig:SchematicSystems} (b)). In this study, we investigate the transition between the 3 metastable states visible in the $\phi_1$, $\phi_2$ descriptor space, namely the configuration close to $\phi_1,\phi_2=1,-1$, $\phi_1,\phi_2=-1,-1$ and $\phi_1,\phi_2=1,1$. We generate training data concatenating ABMD simulations biasing $\phi_1, \phi_2, \phi_3$ with a spring constant of $\kappa=100$~kJ/mol respectively in order to connect the 3 metastable states.
To show the advantages of the selective reconstruction loss introduced in Eq.~\ref{eq:TrainingLossTeacher} we use all 6 dihedral angles as initial features setting $\alpha=0$ for $\phi_3$ and the $\psi$ dihedrals.
To save computational costs we use the FPS tool to reduce the number of training datapoints ($m=1000$) for the teacher model and ($m=1500$) for the student models. The hyperparameters for the teacher model are $k=3$, $t=0.1$, $\beta=1$, $\gamma=1e^{-4}$. After training we use the $s$ variable as a CV to bias a 5 ns long simulation with OPES-EXPLORE. We choose a PACE of 500 steps and a barrier parameter equal to 80 kJ/mol. We apply a harmonic constraint with a cutoff value of 0.2 and $\kappa = 500$~kJ/mol (UWALL) as a function of the $z$ variable. We add the newly sampled datapoints to expand the training dataset and train the DeepLNE student models for $s$ and $z$ setting $\gamma=1e^{-6}$. Then, we proceed by using the student model $s$ variable as a CV to bias 200 ns of MD simulation with OPES in order to estimate the free-energy of the system. We set a PACE of 1000 steps and a barrier parameter equal to 80 kJ/mol. Again, a harmonic constraint with a cutoff value of 0.2 and $\kappa = 5000$~kJ/mol (UWALL) is applied to the $z$ variable. All simulations are carried out \textit{in vacuo} with the DES-Amber ff.~\cite{Piana2020}. We estimate statistical errors on the FES using 10 blocks of length 20 ns.

\subsubsection{Pre-miR21}
\label{MMpremir21}
As a challenging system with a high degree of configurational complexity, we choose the RNA strand pre-miR21. This system consists of 31 nucleotides containing the physiologically relevant Adenosine nucleobase at 5'-3' position 29 (A29) that can perform rotations between a 'stacked-in' state (i.e. A29 pointing towards the paired RNA bases) (see Fig.~\ref{fig:SchematicSystems}) and a 'bulged-out' state, in which A29 is translated in the bulk solution. Recently, pre-miR21 has been studied using extensive enhanced sampling MD simulations~\cite{Aureli2024} using the multi-replica variant of OPES, i.e. OneOPES, where 8 concurrent replicas of the system are simulated under the effect of an increasing thermal ramp. 
This study uses the FES obtained by a simulation with a length of $\sim$400~ns per replica as a reference. Moreover, in the present study, we keep the simulation protocol identical to Ref.~\onlinecite{Aureli2024}, however we perform only a single enhanced sampling MD simulation (using 1 replica).
To sample the transition from state A (A29 'stacked-in') to state B (A29 'bulged-out') we use ABMD simulations biasing 3 different CVs simultaneously: 1) a primary contact map built using the heavy atom distances between the nucleobase A29 and the proximal nucleobases G45, C46 with cutoff 5 and 4 $\angstrom$ respectively (CMAP1); 2) a contact map constructed using the heavy atom distances between the nucleobase A29 and the proximal nucleobases C28, G30 with cutoff 4 $\angstrom$ (CMAP2); 3) a contact map built using the heavy atom distances between the nucleobases C28, G30 with cutoff 8.5 $\angstrom$ (CMAP3). We start multiple ABMD runs from both states at CMAP3 values that span the entire possible range to allow maximum variation in this descriptor space direction, terminating the simulations upon reaching the respective opposite state.
We reduce the number of training datapoints to $m=1000$ via FPS and choose the same distance metrics used for the ABMD simulations as input features in the DeepLNE training. The teacher model is trained with hyperparameters $k=3$, $t=0.1$, $\beta=10$, $\gamma=1e^{-8}$. In the enhanced sampling simulation using the DeepLNE teacher model to expand the training data for the student models, we use the $s$ and $z$ variables to deposit bias with OPES-EXPLORE for a total simulation time of 20 ns. We choose a PACE of 500 steps, bandwidths 0.1 and 1 respectively, and a barrier parameter equal to 60 kJ/mol. We concatenate all available sampled trajectories, subsample a lower number of datapoints via FPS ($m=1500$), and continue by training the DeepLNE student model $s$ setting $\gamma=1e^{-6}$. After completing the training and exporting the student model into PLUMED we use $s$ as a CV to deposit bias via OPES. We set a PACE of 5000 steps and a barrier parameter equal to 35 kJ/mol. With this setup, a trajectory of  2~$\mu s$ has been produced. The free energies were calculated by integrating the probability of being in respective microstates $i$ of the CV by $F = - k_BT \log ( \sum_{i, \text{CV}} p(x_i))$. Statistical errors of the FES were estimated using 10 blocks of 200 ns. For the reference FES, we choose 10 blocks of 23 ns skipping the first 30 ns of the OneOPES simulation.

Simulations of alanine tripeptide and pre-miR21 are performed using the GROMACS 2022.5 engine~\cite{abraham2015gromacs} patched with the PLUMED 2.9 plugin~\cite{Tribello2014} with the Hamiltonian replica exchange algorithm.~\cite{Bussi2014}

\section{\label{sec:Results}Results}

\subsection{Particle in M\"uller-Brown-Potential}
\label{Muller_Brown_Potential}

\begin{figure*}
\includegraphics[width=\textwidth]{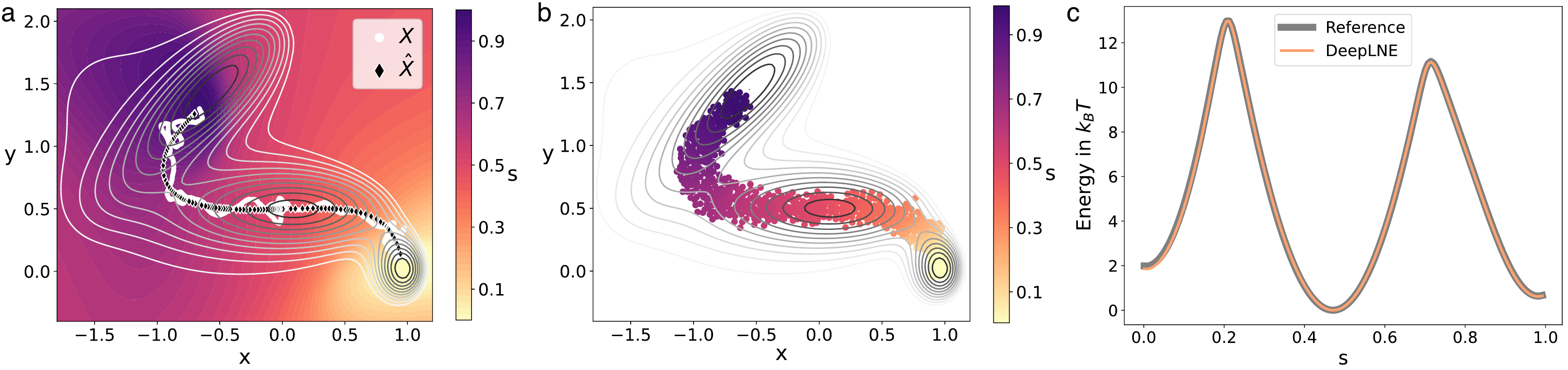}
\caption{\label{fig:MullerBrown} Results of training DeepLNE for a particle simulation in the 3-state M\"uller-Brown-potential.
(a) Training datapoints for DeepLNE, obtained by concatenating 2 ABMD (biasing y-coordinate with a spring constant of $\kappa=30,80$~$k_BT$ respectively. The 2D coordinate space is colored based on the trained DeepLNE $s$ variable. Also we superimpose the $\hat{\bm{X}}$ of the DeepLNE model that is used to compute the $z$ variable.
(b) Sampled configuration during the biased simulation using OPES-EXPLORE applied on the trained PyTorch DeepLNE CVs and a harmonic constraint applied on $z$. The color of the datapoints correspond to the DeepLNE $s$ variable showing path-like behavior. These datapoints are subsequently used to train the DeepLNE student models.
(c) Free energy landscape with respect to the $s$ variable of the student model. We compare the FES obtained via OPES biasing the student model $s$ and applying harmonic restraint on student model $z$ with the FES of an OPES simulation biasing x and y and subsequently reweighting with respect to both DeepLNE student model variables $s$ and $z$.}
\end{figure*}


In Fig.~\ref{fig:MullerBrown} we collected the results of some steps along the proposed DeepLNE knowledge distillation algorithm. In Fig.~SI~1 the remaining intermediate steps can be found. The M\"uller-Brown-Potential used in this study encompasses 3 states. To show the effectiveness of the DeepLNE model we choose a relatively minimal architecture with 1 hidden layer and 3 nodes for the dimensionality reduction $d=3$ and 2 hidden layers with 9 and 3 hidden nodes for both neighborhood compression and decoder network and train it on a concatenation of only 2 ABMD trajectories connecting the respective end states with the intermediate state. After 5000 training epochs we can project the DeepLNE teacher model on the input feature space $\bm{X}$ as shown in Fig.~\ref{fig:MullerBrown} (a). We project the behavior of the $s$ variable of and can appreciate that the choice of low k=3 leads to a DeepLNE model with high locality, which can be seen especially for the regions in phase space with minimal and maximal $s$ values. The smooth extrapolation behavior of the DeepLNE model originates from the continuous k-nearest neighbor step. To achieve a similar extrapolation in the DeepLNE student model we proceed by performing an OPES-EXPLORE simulation biasing $s$ and thereby expanding the training datapoints. The results are shown in Fig.~\ref{fig:MullerBrown} (b) where we find the entire range of $s$ values explored and importantly a broader range of $z$ values is explored too. For both DeepLNE student models $s$ and $z$ we choose an architecture with 3 hidden layers and 8 nodes each. Training the DeepLNE student models on the concatenation of the data $\bm{X}$ shown in Fig.~\ref{fig:MullerBrown} (a) and the new sampled data in (b) and subsequently biasing $s$ and simultaneously applying a harmonic restraint on $z$ we obtain the 1D FES along $s$ shown in Fig.~\ref{fig:MullerBrown} (c). Importantly the estimated free energy is identical to a reweighted trajectory biasing both Cartesian coordinates $x$ and $y$ using the same OPES setting as used when biasing the path-like DeepLNE CVs.
In Fig.~SI~1 (a) we give additional insight into the labels assigned for the supervised multi-tasking framework of DeepLNE. Each datapoint in known regions of the descriptor space is labeled allowing for tailoring the directionality learned by the model during training.
In Fig.~SI~1 (b) we projected the DeepLNE variable $z$ on the descriptor space as done for $s$ in Fig.~\ref{fig:MullerBrown} (a). As expected we find the reconstructed data $\bm{\hat{X}}$ in the center of the training datapoints and accordingly the computed $z$ variable describes a reaction channel connecting all 3 system states.
In Fig.~SI~1 (d), we go beyond the 1D projection of the FES and use the additional information available from the $z$ variable to create the 2D FES obtained from the same enhanced sampling simulation run that resulted in Fig.~\ref{fig:MullerBrown} (c). The 2D FES projection allows for additional \textit{a posteriori} analysis revealing transition regions and more detailed location as well as the extend of the metastable states. This will be especially useful when attempting to describe systems of increasing degree of complexity.

\subsection{Alanine Tripeptide}
\label{Alanine_dipeptide}

\begin{figure*}
\includegraphics[width=\textwidth]{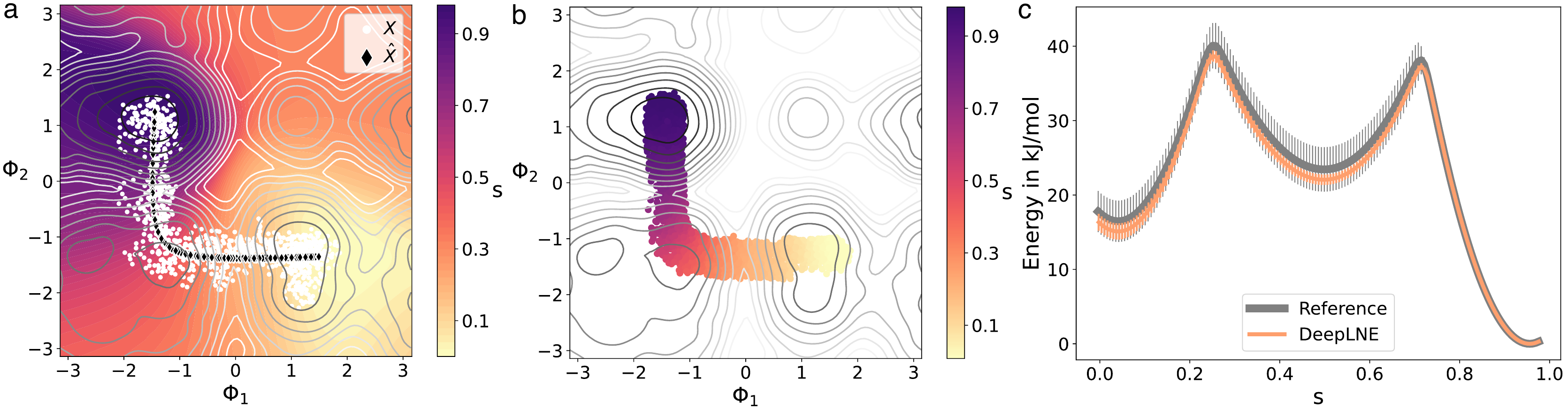}
\caption{\label{fig:Trialanine} Results of training DeepLNE for alanine tripeptide.
(a) Training datapoints for DeepLNE, obtained by concatenating ABMD (biasing $\phi1, \phi2, \phi3$ with a spring constant of $\kappa=100$~kJ/mol respectively. The 2D coordinate space with respect to $\phi1$ and $\phi2$ is colored based on the trained DeepLNE $s$ variable (we set all remaining $\phi$ and $\psi$ input variables to $0$). Also we superimpose the $\hat{\bm{X}}$ of the DeepLNE model.
(b) Sampled configuration during the biased simulation using OPES-EXPLORE applied on the trained PyTorch DeepLNE CVs and a harmonic constraint applied on $z$. The color of the datapoints correspond to the DeepLNE $s$ variable showing path-like behavior. These datapoints are subsequently used to train the DeepLNE student models.
(c) Free energy landscape with respect to the student model $s$ variable for an OPES biased simulation using the student model $s$ and harmonic restraint on student model $z$ variable as well as an OPES simulation biasing $\phi1$ and $\phi2$ and the same harmonic restraint on student model $z$ variable.}
\end{figure*}


The alanine tripeptide system has been widely studied showing that it can be sufficiently described via its 6 dihedral angles $\phi$ and $\psi$.  In this study, even though alanine tripeptide has more states that constitute the entire configurational space, we treat it as a 3 state system considering the configurations $\phi_1,\phi_2=1,-1$, $\phi_1,\phi_2=-1,-1$ and $\phi_1,\phi_2=-1,1$ as the metastable states. Additionally we consider a specific transition path that the system has to follow. In Fig.~\ref{fig:Trialanine} (a) we show the training data obtained from multiple ABMD simulations describing this reaction path. The trajectories are connecting all 3 states and we proceed by training a DeepLNE teacher model. The trained model takes as input all 6 $\phi$ and $\psi$ angles and its $s$ variable is projected in along $\phi_1$ and $\phi_2$ by setting the $\psi$ angles to 0 and $\phi_3=-1$. We choose an architecture with 1 hidden layer and 3 nodes for the dimensionality reduction $d=3$ and 1 hidden layers with 2 hidden nodes for both neighborhood compression and decoder network. Also we select k=3 to obtain a DeepLNE model with high locality, which is reflected in the regions in dihedral space with minimal and maximal $s$ value (compare Fig.~\ref{fig:Trialanine} (a)). The DeepLNE teacher is capable of smooth extrapolation originating from the continuous k-nearest neighbor step. We proceed by expanding the training data coupling the teacher model CV to OPES-EXPLORE simulation biasing $s$ and applying a harmonic restraint on $z$ in order to exclusively sample the envisioned reaction path. The results are shown in Fig.~\ref{fig:Trialanine} (b) where we find the entire range of $s$ values explored and importantly obtain a more densely sampled descriptor space $\bm{X}$ along the path. We train the DeepLNE student models on the concatenation of the data shown in Fig.~\ref{fig:Trialanine} (a) and the newly sampled data in (b). Because of the functional complexity differences between $s$ and $z$ in this system the DeepLNE $s$ variable can be approximated with ANNs of lower complexity compared to the DeepLNE $z$ variable. Accordingly for the student model $s$ we choose a smaller architectures with 1 hidden layer and 3 nodes compared to the $z$ student model with 2 hidden layers and 6 and 3 nodes each.
We can appreciate that the training datapoints describe the transition regions between the metastable states at $\phi_1,\phi_2=0,-1$ and $\phi_1,\phi_2=-1,0$ respectively. This is a specific choice for the transition path, but not the only relevant one considering the entire available configurational space for alanine tripeptide. Since dihedral angles are periodic the 3 metastable states that the $s$ variable connects can be reached by crossing also other transition regions visible in (a) and (b). This has the consequence that when we want to compare FES estimations obtained from DeepLNE biased simulations to a reference simulation that can sample the entire configurational space, we need to restrict the latter such that both enhanced sampling simulations sample the same reaction path, which can be achieved by applying harmonic restraints on $z$ in both cases. Accordingly we proceed by training the student models for $s$ and $z$ on the expanded training dataset and perform an OPES simulation biasing $s$ and restraining the system along $z$ to stay within the reaction tube constituted by the variables $s$ and $z$. In Fig.~\ref{fig:Trialanine} (c) we then compare the FES obtained from a DeepLNE biased run with a reference run biasing all 3 $\phi$ dihedral angles. We can appreciate that both simulations arrive at the same estimation of the FES in the reaction channel subspace defined via the DeepLNE student model variable $z$.
In Fig.~SI~2 (a) we detail how the training data were labeled in order to supervise DeepLNE while constructing the path-like variables $s$ and $z$. Each datapoint in the 3 states within the descriptor space is labeled allowing for the desired directionality of $s$.
In Fig.~SI~2 (b) we projected the DeepLNE variable $z$ on the descriptor space as done for $s$ in Fig.~\ref{fig:Trialanine} (a). The reconstructed data $\bm{\hat{X}}$ and the corresponding $z$ variable that is a function of $\bm{\hat{X}}$ describes a reaction channel connecting all 3 system states that is a subspace of the entire configurational space. In difference to the toy model investigated in \ref{Muller_Brown_Potential} there are states with significant probabilistic weight such that comparison to reference simulations need to be performed within the same reaction channel created by $s$ and $z$. Respecting this condition the FES can be estimated in the 2 DeepLNE dimensions $s$ and $z$ as shown in Fig.~SI~2 (d). We can appreciate that all metastable states as well as the transition regions appear as regions of orthogonal orientation with respect to the $s$ variable indicating that the created reaction channel by the combination of $s$ and $z$ crosses the transition states with high orthogonality, which is indicative of a CV that is approximating well the true reaction coordinate of the system (compare committor analysis performed in \cite{Froehlking2024}).

\subsection{Pre-miR21}
\label{pre-miR21}
\begin{figure*}
\includegraphics[width=\textwidth]{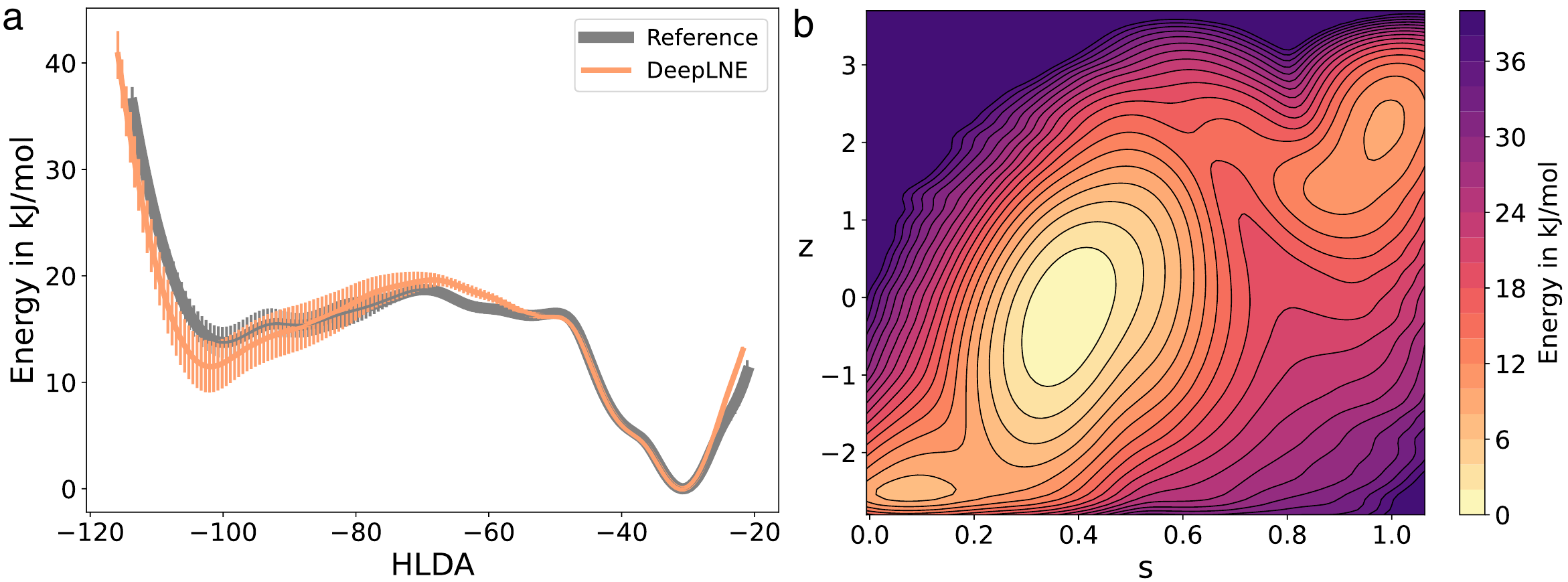}
\caption{\label{fig:RNA} Results of training DeepLNE for pre-miR21.
(a) Free energy landscape with respect to the HLDA variable used by Ref.~\onlinecite{Aureli2024} comparing to a OPES simulation using the student model $s$ to deposit bias. The nucleobase A29 is 'stacked-in' the helix at high values of HLDA and 'bulged-out' towards the solvent at low HLDA values.
The reference study used OneOPES with 8 replicas and we evaluate the FES obtained after a total simulation time of more than 400~ns per replica. In this study we perform a single OPES simulation instead biasing the student model $s$ variable for 2 $\mu s$. 
(b) Free energy landscape with respect to the student model $s$ and $z$ variable showing 3 metastable states of the nucleobase A29 being 'stacked-in' at values of $s$ below 0.7 and 'bulged-out' at higher $s$ values (compare Fig.~SI~4 for the centroids associated to the 3 metastable states identified here). This description of the system along the DeepLNE variables also reveals the transition state regions along the reaction path.}
\end{figure*}


The microRNA pre-miR21 is a system of high configurational complexity, namely the number of available descriptors, significant degrees of freedom, and metastable states is difficult to define \textit{a priori}. While the challenge of reducing the descriptor dimensionality can be addressed using DeepLNE by transforming the high dimensional input feature space into a single CV $s$, the latter challenge can only be overcome by identifying a sub-set of the whole configurational space. Pre-miR21 has been recently studied with respect to processes for which base-flipping of the A29 is of relevance \cite{Aureli2024}. Following the previous paper, the current study focuses on the same configurational re-arrangement. We note that this is an important first step that decides all the following analysis and results. 
As described in section \ref{MMpremir21}, we selected three CMAPS as input features for the DeepLNE teacher model. These CMAPS describe both the distance between A29 and its neighboring nucleobases in the base-paired state and the distance between the stacked nucleobases above and below A29.
Multiple ABMD simulations make up the training data distributed maximally diverse in the descriptor space. As shown in Fig.~SI~3 (a), we make use of the multi-tasking framework assigning labels to both end states known from the previous study as well as an \textit{ad hoc} defined intermediate that is located between the end states when projecting the states in the input feature space $\bm{X}$. We choose an architecture with 1 hidden layer and 3 nodes for the dimensionality reduction $d=3$ and 1 hidden layer with 3 nodes for both neighborhood compression and decoder network. Also, we select k=3 to obtain a DeepLNE model with high locality. We proceed by training the DeepLNE teacher on the labeled training data, exporting it into PLUMED, and performing OPES-EXPLORE simulation biasing exclusively the $s$ variable. Combining the new datapoints (see Fig.~SI~3 for (b) DeepLNE $s$ variable and (c) DeepLNE $z$ variable) sampled with the already available ABMD trajectories the DeepLNE student models are trained on reconstructing the projection of the DeepLNE teacher on those data. As demonstrated in alanine tripeptide we choose a smaller architecture with 1 hidden layer and 3 nodes for the student model $s$ and a slightly more flexible architecture for the $z$ student model with 2 hidden layers and 3 nodes each. In Fig.~\ref{fig:RNA} we collected the results of the OPES simulation performed using the trained DeepLNE student model biasing the variable $s$. In Fig.~\ref{fig:RNA} (a) we compare the FES obtained in this study with the one previously obtained in Ref.~\onlinecite{Aureli2024} biasing a harmonic linear discriminant analysis (HLDA) CV (i.e., a weighted linear combination of 10 inter-nucleobases distances). We project the FES obtained from DeepLNE biased simulation along the same CV showing that it converges to the same FES estimation. We note that while the previous study used a total simulation time of over 3 $\mu s$ using multiple replicas we obtained the same result after 2 $\mu s$ in a single MD realization. Additionally in Fig.~\ref{fig:RNA} (b) we investigate the underlying dynamical rearrangement of A29 in more detail by projecting the FES along both DeepLNE student model variables $s$ and $z$. We can appreciate that while the HLDA CV used in the previous study can distinguish the 2 metastable end states 'stacked-in' and 'bulged-out', the DeepLNE variables $s$ and $z$ reveal 3 metastable states as well as the interconnecting transition regions. In Fig.~SI~4 we extracted the centroids of the structural ensemble associated with the 3 metastable states that can be identified in Fig.~\ref{fig:RNA} (b). While the end states are the previously identified conformations with A29 being 'stacked-in' ($s$<0.2) or 'bulged-out' respectively ($s$>0.9) and the intermediate metastable state corresponds to the approaching of A29 into the RNA helix, with A29 starting to form the necessary contacts to achieve the fully 'stacked-in' state.

\subsection{Performance Benchmarks}

To quantify the computational speed-up that can be achieved by applying the knowledge distillation framework introduced in this study we compare the MD simulation performance when biasing the DeepLNE teacher model with DeepLNE student models of increasing number of parameters. The performances of all the models are evaluated using an AMD Ryzen 7950X CPU and a Nvidia RTX3080 GPU. Errors are estimated on 3 repetitions of each benchmark simulation. In Fig.~\ref{fig:RNA} we collected the results of the average DeepLNE student model performance benchmark measure in ns/day for alanine tripeptide (42 atoms) and the solvated RNA pre-miR21 (45543 atoms). The displayed values should be compared to the respective PyTorch-based CV teacher model performance of 201 ns/day for alanine tripeptide and 119 ns/day for pre-miR21 corresponding to speed-up factors of up to 20 and 3 respectively when using the student models. Importantly we can appreciate that across more than 4 orders of magnitude in student model complexity the performance remains similar. While a student model for alanine tripeptide with 16 parameters leads to an average performance 3959 ns/day and a more complex model with 20000 parameters still achieves 3774 ns/day. A student model for pre-miR21  with 16 parameters results in an average performance 329 ns/day while a model with 20000 parameters arrives at 325 ns/day.
These results show that DeepLNE student models offer the potential to approximate and bias functions of increasing complexity while achieving similar MD performance.




\begin{figure*}[htb]
\includegraphics[width=0.5\textwidth]{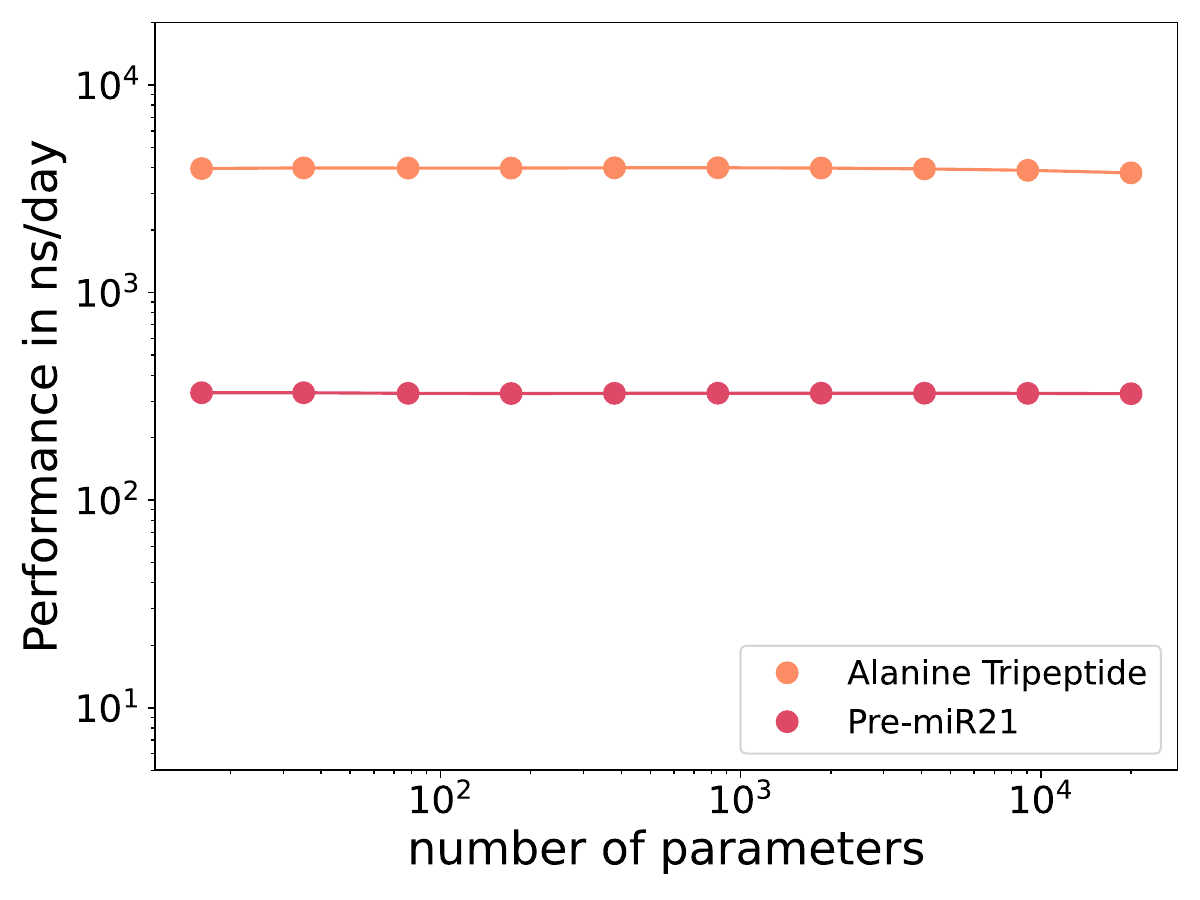}
\caption{\label{fig:RNAPerformance} Computational performance of DeepLNE student models s comparing increasing complexity of the architectures, resulting in increasing number of model parameters. For alanine tripeptide \textit{in vacuo} and the pre-miR21 RNA system investigated the performance measure in ns of simulation time per day remains constant with respect to increased model complexities. As reference, we note that the corresponding PyTorch teacher model for alanine tripeptide allows 201 ns/day while for pre-miR21 119 ns/day can be achieved. Errors are increasing with increasing model parameters where maximal errors on the performance estimates are 388 ns/day for alanine tripeptide and 6 ns/day for pre-miR21.}
\end{figure*}

\section{\label{sec:Discussion}Discussion}

In this study, we propose the DeepLNE++ strategy to construct a computationally fast DeepLNE CV using a knowledge distillation approach. The algorithm selects the most generalizing model at each step, supervising its directionality via the assignment of labels to the training data and representing the path-like DeepLNE variables $s$ and $z$ using ANN student models. The new algorithm can be used to capture and especially accelerate rare events in large biophysical systems with reduced computational overhead. In multiple systems, we have shown how this procedure allows exporting computationally efficient CVs in MD software and maintaining its capability to approximate well the ideal reaction coordinate. Consequently, we could demonstrate that the DeepLNE++ CVs can be effectively used to improve sampling when coupled to enhanced sampling methods such as OPES without multi-replica approaches.

Our methodology starts by obtaining reactive trajectories that cover the descriptor space in the most diverse manner possible encompassing the transitions between all metastable system states of interest. Herein, these starting trajectories have been obtained with a ratchet-and-pawl restraint~\cite{Camilloni2011}, which provided reactive data of sufficient quality to successfully train DeepLNE models for all the studied systems. 
The input trajectory is used to define a feature vector to distinguish all states encountered and their intermediates. With this featurized trajectory, one can automatically derive a path-like DeepLNE teacher model. This model has advantageous extrapolation capabilities because it is intrinsically anchored to the training data. To maximally maintain this behavior when approximating the teacher model using ANNs as DeepLNE student models, we rely on an explorative enhanced sampling step, where we use the DeepLNE teacher model to bias the system dynamics in order to expand the training dataset. The interpretation of the derived variables $s$ and $z$ is identical in DeepLNE teacher and student models, as they describe the progress along the path and its perpendicular distance, respectively. 

The proposed method achieves a speed-up in computational model evaluation when biasing MD simulations with DeepLNE. It does so by representing the DeepLNE model containing a large number of fitting parameters with ANNs of lower complexity. The DeepLNE++ approach uses cross-validation to choose the models with the highest robustness when extrapolating into unknown regions of the descriptor spaces. Furthermore, we represented the DeepLNE variables $s$ and $z$ by an ANN each, instead of using a single ANN to reconstruct both variables in the output. This decision is made because one would need to normalize the outputs and also ensure for each system that the objective function is leading to equally distributing loss minimization for each variable simultaneously.

We keep some hyperparameter choices to the user as regularization options that permit tailoring a path-like CV adapt to system-specific needs. The model training steps of the algorithm are computationally fast and can additionally employ available GPUs and therefore allow fine-tuning all hyperparameter choices of DeepLNE leading to CVs of high quality.
We demonstrated how the hyperparameters within the DeepLNE knowledge distillation framework allow for regulating the models. For all the DeepLNE models we tested $k=3$ and $t=0.1$ in order to establish a continuous k-nearest-neighbor selection step that allows to obtain DeepLNE teacher models with high locality and numerical stability. The utility of the hyperparameter $\alpha$ was shown for alanine tripeptide where the relevant dihedral angle input descriptors could be selected to describe the desired transition path. Although not implemented in this study, the selection of $\alpha$ for each feature can be automated by adding an additional L1-norm or Softmax term for $\alpha$ to the objective function. This would allow for automatic assignment of feature importance weights. In all 3 toy models, it was necessary to assign a significantly large value to the hyperparameter $\beta$ in order to supervise DeepLNE with respect to the directionality and sequentiality of the variables $s$ and $z$. The hyperparameter $\gamma$ that is penalizing the magnitude of the fitting parameters in teacher as well as student models is chosen relatively low in all 3 toy models. This resulted in models of higher functional flexibility. However, because we were choosing the models that minimize the cross-validation error in all cases the models exhibited sufficiently smooth extrapolation behavior and numerical stability when applied to the task of CV-based free energy estimation using enhanced sampling MD.

We notice that our proposed knowledge distillation framework is not limited to the representation of DeepLNE CVs as computationally fast ANNs, but can be used to approximate any computationally expensive CV.

We show that our method can be straightforwardly applied to multi-state systems of biologically relevant complexity at a fraction of the computational cost of the original DeepLNE CV \cite{Froehlking2024}. A DeepLNE student model can be successfully trained as long as the training trajectories visit the entire descriptor space encompassed by the reaction channel defined by the variables $s$ and $z$. In the systems studied here, we impose a sequentiality of the metastable states into the DeepLNE $s$ variable in order to show that path-like CVs for multi-state systems benefit from a supervised multi-tasking objective function during the training step. 
In summary, we believe that the advances presented in this study significantly broaden the applicability of DeepLNE++, especially for large and complex biological systems. The introduction of automatic, efficient, and fast path-like CVs in DeepLNE++ is crucial for accelerating MD sampling of their dynamics and reactions.

\section*{Supplementary Material}
Sections for each toy model showing 
(1) M\"uller-Brown-Potential with additional details of the DeepLNE training steps and resulting 2D FES of the biased MD simulations, 
(2) Alanine Tripeptide details of the DeepLNE training steps and resulting 2D FES of the biased MD simulations, 
(3) Pre-miR21 training datapoints for the DeepLNE models and structural analysis of the identified metastable states from the enhanced sampling MD simulations.

\section*{Acknowledgments}
We acknowledge Luigi Bonati for useful discussion and providing many suggestions. 
F.L.G., V.R., S.A. and T.F. acknowledge the Swiss National Supercomputing Centre (CSCS) for large supercomputer time allocations projectID:s1228. They also acknowledge the Swiss National Science Foundation and Bridge for financial support (projects numbers: $200021\_204795$, $CRSII5\_216587$ and $40B2-0\_203628$).

\section*{Data availability}
At \url{https://github.com/ThorbenF/DeepLNE2} we provided tutorials together with scripts that allow to reproduce the figures and the results of this study.

\bibliography{main}

\clearpage
\onecolumngrid
\appendix 

\appendix

\renewcommand{\thesection}{SI \arabic{section}}
\newcommand{\SIsection}[1]{\refstepcounter{section}\section*{\thesection. #1}\addcontentsline{toc}{section}{\thesection. #1}}

\renewcommand{\thefigure}{SI \arabic{section}.\arabic{figure}}

\makeatletter
\renewcommand{\fnum@figure}{\figurename~\thefigure}
\renewcommand{\@makecaption}[2]{%
  \vskip\abovecaptionskip
  \sbox\@tempboxa{{\bfseries SI FIG \thefigure.} #2}%
  \ifdim \wd\@tempboxa >\hsize
    {FIG. SI \thefigure.} #2\par
  \else
    \global \@minipagefalse
    \hb@xt@\hsize{\hfil\box\@tempboxa\hfil}%
  \fi
  \vskip\belowcaptionskip}
\makeatother

\counterwithout{figure}{section}
\setcounter{figure}{0}

\begin{center}
    {\large \bfseries Supplementary Material - DeepLNE++ leveraging knowledge distillation for accelerated multi-state path-like collective variables \par}
    \vspace{1em}
    {\normalsize
    Thorben Fr\"ohlking\textsuperscript{1,2,3}, Valerio Rizzi\textsuperscript{1,2,3}, Simone Aureli\textsuperscript{1,2,3}, Francesco Luigi Gervasio\textsuperscript{1,2,3,4}\par}
    \vspace{1em}
    {\small
    \textsuperscript{1}School of Pharmaceutical Sciences, University of Geneva, Rue Michel Servet 1, 1206, Genève, Switzerland \par
    \textsuperscript{2}Institute of Pharmaceutical Sciences of Western Switzerland (ISPSO), University of Geneva, 1206, Genève, Switzerland \par
    \textsuperscript{3}Swiss Institute of Bioinformatics, University of Geneva, 1206, Genève, Switzerland \par
    \textsuperscript{4}Department of Chemistry, University College London, London, WC1E 6BT, United Kingdom \par
    \vspace{1em}
    \texttt{francesco.gervasio@unige.ch} \par
    }
    \vspace{2em} 
\end{center}

\SIsection{Particle in 3-state M\"uller-Brown-Potential}
\begin{figure}[H]
\centering
\includegraphics[width=0.95\textwidth]{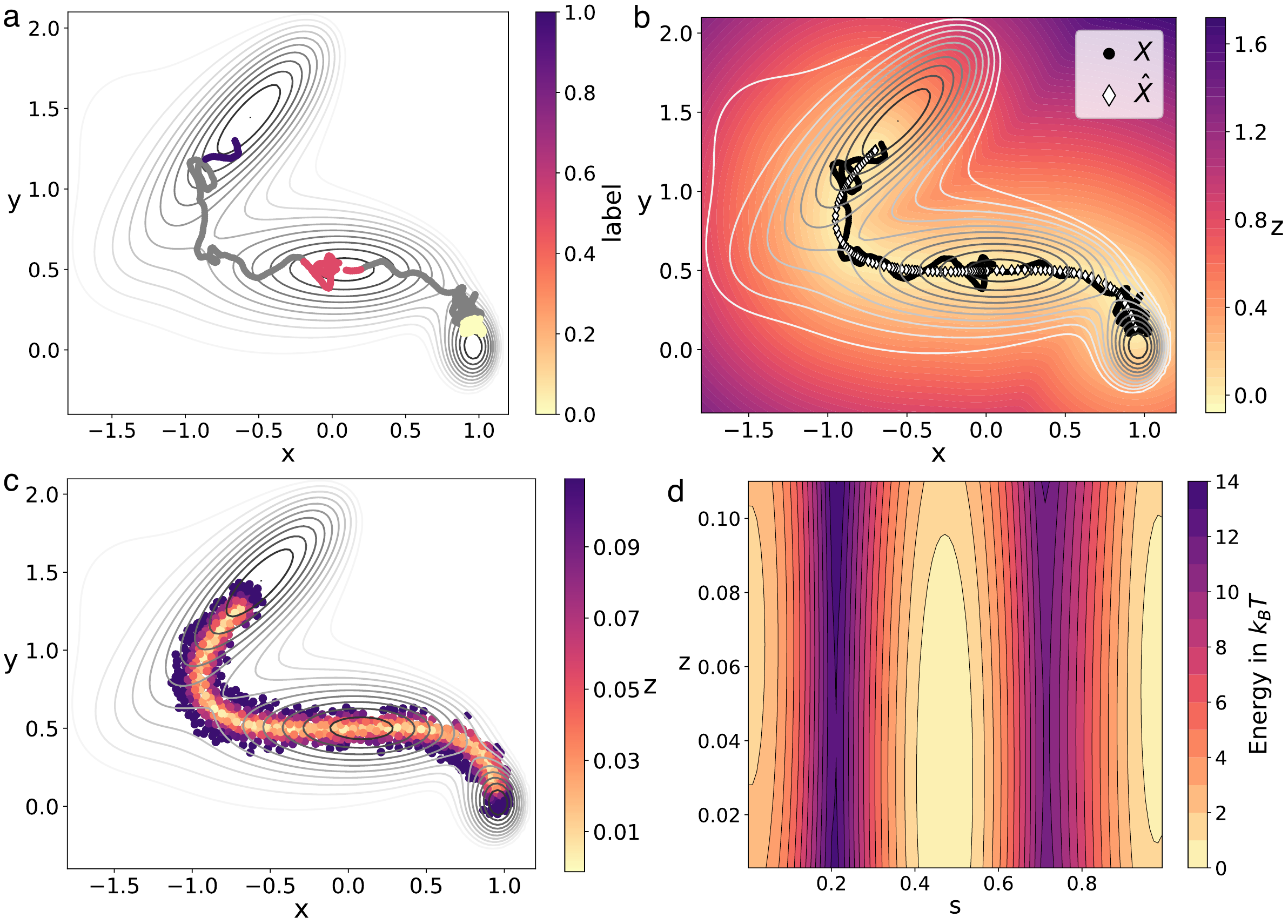}
\caption{\label{AppendixFig:MullerBrown} Results of training DeepLNE for a particle simulation in the 3-state M\"uller-Brown-potential.
(a) Training datapoints for DeepLNE, obtained by concatenating 2 ABMD (biasing y-coordinate with a spring constant of $\kappa=30,80$~$k_BT$ respectively. The datapoints are colored based on the label for the 3 states (0, 0.5, 1).
(b) Training datapoints superimposed with the 2D coordinate space is colored based on the trained DeepLNE $z$ variable with a $\lambda=10$. Also we superimpose the $\hat{\bm{X}}$ of the DeepLNE model.
(c) Sampled configuration during the biased simulation using OPES-EXPLORE applied on the trained PyTorch DeepLNE CVs and a harmonic constraint applied on $z$. The color of the datapoints correspond to the DeepLNE $z$ variable showing path-like behavior. These datapoints are subsequently used to train the DeepLNE student models.
(d) Two-dimensional free energy landscape with respect to the DeepLNE $s$ and $z$ variable obtained from biased MD using OPES on the student model $s$ and harmonic restraint on student $z$. This is especially useful when one is interested in identifying the transition states in more detail.}
\end{figure}


\SIsection{Alanine Tripeptide}
\begin{figure}[H]
\includegraphics[width=\textwidth]{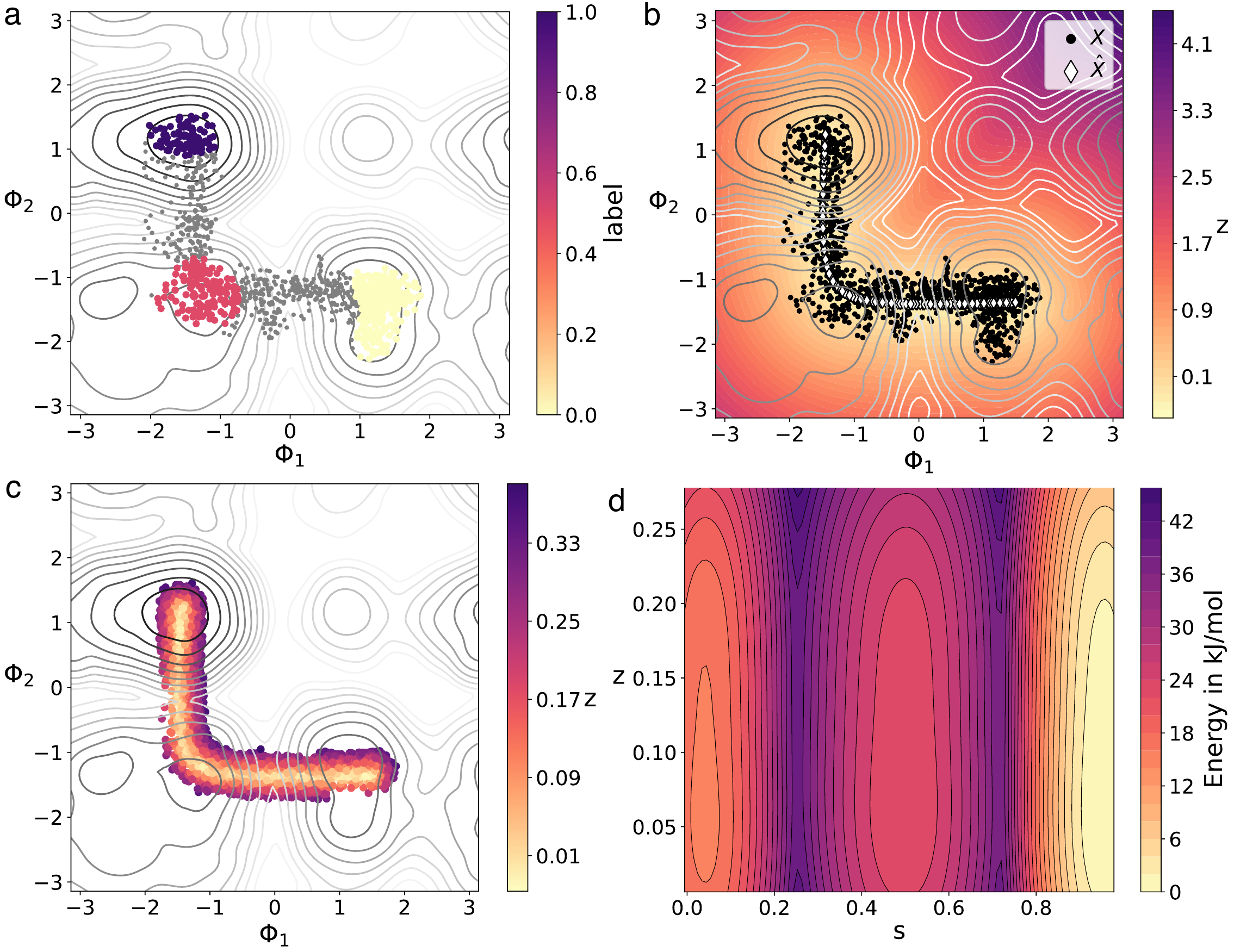}
\caption{\label{AppendixFig:Trialanine} Results of training DeepLNE for a Alanine Tripeptide simulation.
(a) Training datapoints for DeepLNE, obtained by concatenating ABMD (biasing $\phi1, \phi2, \phi3$ with a spring constant of $\kappa=100$~kJ/mol respectively. The datapoints are colored based on the label for the 3 states (0, 0.5, 1).
(b) Training datapoints for DeepLNE superimposed with the 2D coordinate space colored based on the trained DeepLNE $z$ variable with a $\lambda=10$ (we set all remaining $\phi$ and $\psi$ input variables to $0$). Also we superimpose the $\hat{\bm{X}}$ of the DeepLNE model. 
(c) Sampled configuration during the biased simulation using OPES-EXPLORE applied on the trained PyTorch DeepLNE CVs and a harmonic constraint applied on $z$. The color of the datapoints correspond to the DeepLNE $z$ variable showing path-like behavior. These datapoints are subsequently used to train the DeepLNE student models.
(d) Two-dimensional free energy landscape with respect to the DeepLNE $s$ and $z$ variable obtained from biased MD using OPES on the student model s and harmonic restraint on student $z$. One can identify sharply the transition regions between the 3 metastable states near $s=0.2, z=0.04$ and $s=0.7, z=0.04$.}
\end{figure}


\SIsection{Pre-miR-21}
\begin{figure}[H]
\includegraphics[width=\textwidth]{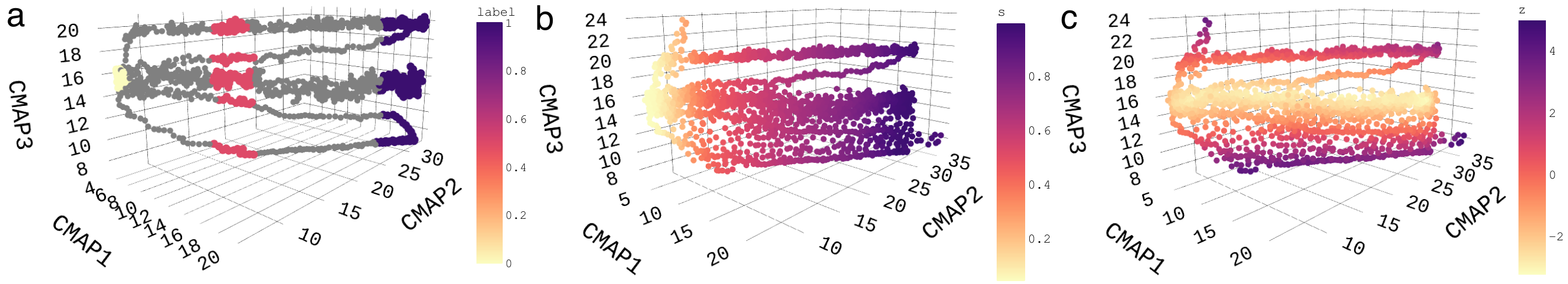}
\caption{\label{AppendixFig:RNA} Results of training DeepLNE for a RNA pre-miR21 simulation.
(a) Training datapoints for DeepLNE, obtained by concatenating 5 ABMD (biasing the 3 CMAP CVs with a spring constant of $\kappa=100$~kJ/mol respectively. The datapoints are colored based on the label for the 3 states (0, 0.5, 1).
(b) Sampled configuration during the biased simulation using OPES-EXPLORE applied on the trained PyTorch DeepLNE CV $s$. The color of the datapoints correspond to the DeepLNE $s$ variable showing path-like behavior. These datapoints are subsequently used to train the DeepLNE student models.
(c) Sampled configuration during the biased simulation using OPES-EXPLORE applied on the trained PyTorch DeepLNE CVs and a harmonic constraint applied on $z$. The color of the datapoints correspond to the DeepLNE $z$ variable with a $\lambda=1$ showing path-like behavior. These datapoints are subsequently used to train the DeepLNE student models.}
\end{figure}

\begin{figure}[H]
\includegraphics[width=\textwidth]{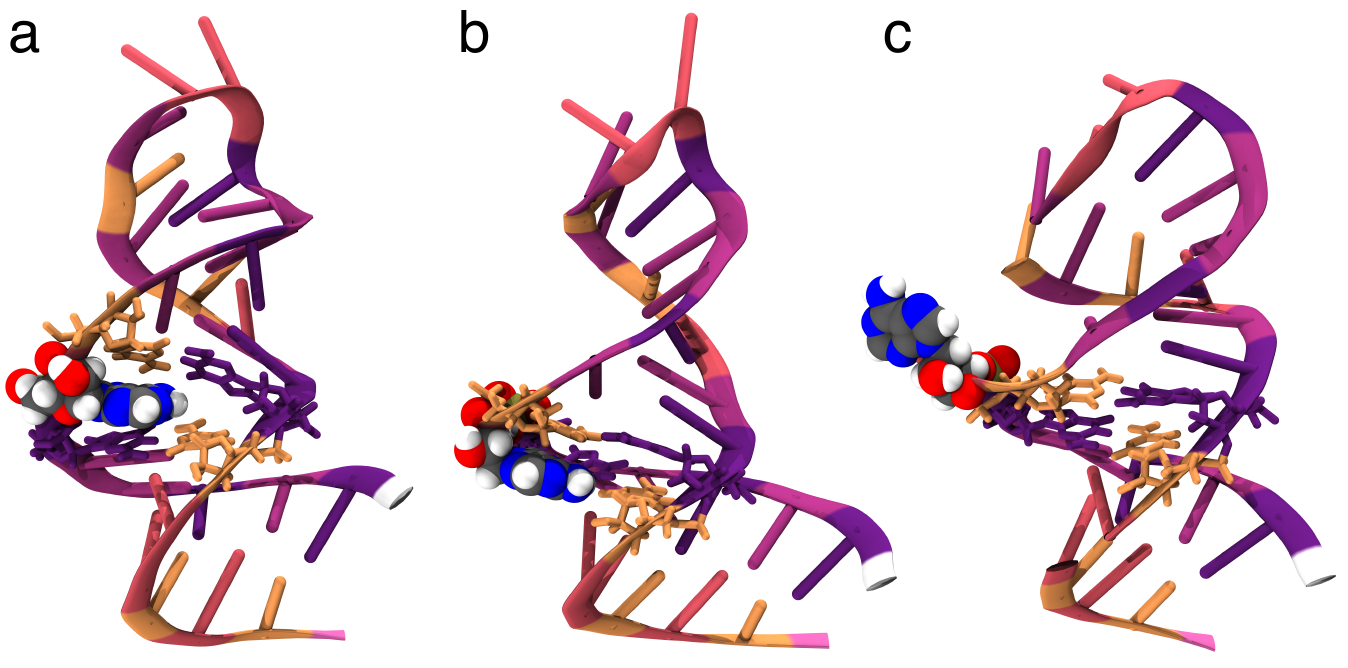}
\caption{\label{AppendixFig:RNA_StructuralAnalysis} Results of structural analysis of the FES for the DeepLNE biased RNA pre-miR21 simulation. The structure of pre-miR21 is composed of 31 nucleotides. The RNA from G22 to C54 is coloured with respect to the type of nucleotide: Guanine-purple, Adenine-light crimson, Cytosine-orange, Uracil-pink, G5-white, C3-mauve. We show the centroids of the 3 minima in the FES shown in the main text. The nucleobase A29 is highlighted via "VDW' and the surrounding nucleobases are represented via 'licorice' style.
(a) State A at $s=0, z=-2.5$, corresponding to the A29 being 'stacked-in'. The nucleobase is integrated within the helical structure forming hydrogen bonds with the surrounding nucleobases.
(b) State B at $s=0.4, z=0$, corresponding to the A29 being 'partially rotated outward'.
(c) State C at $s=1.0, z=2$, corresponding to the A29 being 'bulged-out'.}
\end{figure}

\end{document}